\documentclass[prl,twocolumn,showpacs,aps,floats,floatfix]{revtex4}

\usepackage{rotating} 
\usepackage{graphicx}
\usepackage{tabularx}
\usepackage{axodraw}

\input colordvi.tex

\newcommand{\beq}{\begin{equation}} 
\newcommand{\eeq}{\end{equation}} 
\newcommand{\beqa}{\begin{eqnarray}} 
\newcommand{\eeqa}{\end{eqnarray}} 
\newcommand{\mx}{\left[\begin{array}} 
\newcommand{\finmx}{\end{array}\right]} 
\newcommand{\mxp}{\left(\begin{array}} 
\newcommand{\finmxp}{\end{array}\right)} 
\newcommand{\casos}{\left\{\begin{array}} 
\newcommand{\fincasos}{\end{array}\right.} 
\newcommand{\rcasos}{\left.\begin{array}} 
\newcommand{\rfincasos}{\end{array}\right\}} 

\def\lsim{\ \rlap{\raise 3pt \hbox{$<$}}{\lower 3pt \hbox{$\sim$}}\ }
\def\gsim{\ \rlap{\raise 3pt \hbox{$>$}}{\lower 3pt \hbox{$\sim$}}\ }

\def\epja#1{Eur.~Phys.~J.\ {\bf A#1}}

\def\ea{{\it et al.}}

\begin{document}

 \title{   
 \preprint{astro-ph/0306384}
 \preprint{UdeA-PE/03-09}
Exploring the sub-eV neutrino mass range with supernova neutrinos
}

\author{Enrico Nardi$^{a,b}$ and Jorge I. Zuluaga$^b$} 

\affiliation{
$^a$INFN, Laboratori Nazionali di Frascati, C.P. 13, I00044 Frascati,  Italy \\
$^b$Instituto de F\'\i sica,  Universidad de Antioquia, A.A.{\it 1226}, Medell\'\i n, Colombia}

                               \date{\today}

\begin{abstract}
  A new method to study the effects of neutrino masses on a supernova neutrino
  signal is proposed.  The method relies exclusively on the analysis of the
  full statistics of neutrino events, it is independent of astrophysical
  assumptions, and does not require the observation of any additional
  phenomenon to trace possible delays in the neutrino arrival times.  The
  sensitivity of the method to the sub-eV neutrino mass range, defined as the
  capability of disentangling at 95\% c.l. the case $m_\nu=1\,$eV from $m_\nu=0$,
  is tested by analyzing a set of synthetic neutrino samples modeled according
  to the signal that could be detected at SuperKamiokande.  For a supernova at
  the Galactic center success is achieved in more than 50\% of the cases.  It
  is argued that a future Galactic supernova yielding several thousands of
  inverse $\beta$ decays might provide enough information to explore a neutrino
  mass range somewhat below $1\,$eV.
\end{abstract} 

\bigskip



\maketitle

\newpage



During the past few years, the large amount of data collected by solar
\cite{solar} and atmospheric \cite{atmospheric} neutrino experiments provided
strong evidences for non vanishing neutrino masses.  The recent KamLAND result
\cite{kamland} on the depletion of the neutrino flux from nuclear power plants
in Japan, gave a final confirmation of this picture. Since in the Standard
Model (SM) of particle physics all neutrino species are massless, this
constitutes the first direct evidence for new physics, and provides important
informations for developing theories beyond the SM.  However, to date all the
evidences for neutrino masses come from oscillation experiments, that are only
sensitive to mass square differences and cannot give any informations on
single mass values.  The importance of measuring the absolute value of
neutrino masses cannot be understated. It is presently being addressed by
means of a remarkably large number of different approaches, ranging from
laboratory experiments to a plethora of methods that relay on cosmological
considerations. Recent reviews can be found in \cite{absolutenumass}.  From
the study of the end-point of the electron spectrum in tritium $\beta$-decay,
laboratory experiments have been able to set the limit $m_{\nu_e}< 2.2\,$eV
\cite{tritiumlimit}. This is already close to the sensitivity limit of
on-going experiments.  If neutrinos are Majorana particles, the non
observation of neutrinoless double $\beta$ decay can constrain a particular
combination of the three neutrino masses.  Interpretation of these
experimental results is difficult due to large theoretical uncertainties
related to nuclear matrix elements calculations. This is reflected in a model
dependent limit $m_\nu^{eff} < 0.35 -1.24\,$eV \cite{absolutenumass,doublebeta}.
A tight bound $\sum_i m_{\nu_i} < 0.7\,$eV was recently set by the WMAP
Collaboration \cite{WMAP} by combining measurements of cosmic microwave
background anisotropies with data from bright galaxies redshift surveys
\cite{Colless:2001gk} and other cosmological data.  However, this limit
becomes much looser if the set of assumptions on which it relies is relaxed
(see \cite{Hannestad} for discussions of this point). Therefore it is quite
important to keep looking for alternative ways to measure the neutrino masses.
Model independent limits can be more reliably established by combining
complementary experimental informations and, needless to say, it would be of
utmost importance if the mass values could be measured by means of
more than one independent method.

Already long time ago it was realized that Supernova (SN) neutrinos can
provide valuable informations on the neutrino masses \cite{Zatsepin:1968}.
The basic idea relies on the time-of-flight delay $\Delta t$ that a neutrino of
mass $m_\nu$ and energy $E_\nu$ traveling a distance $L$ would suffer with respect
to a massless particle:
\beq \label{delay}
\frac{\Delta t}{L} = \frac{1}{v} - 1 
\approx 
\left(
\frac{5.1\,\rm ms}{10\,\rm kpc}\right)  
\! \left(\frac{10\, \rm
    MeV}{E_\nu}\right)^2  \!\! 
  \! \left(\frac{m_\nu}{1\,\rm eV}\right)^2\,,
\eeq
where for ultra-relativistic neutrinos we have used $1/v=E_\nu/p_\nu \simeq 1+
m^2_\nu/2 E_\nu^2$.  The dispersion in the arrival time of about twenty electron
anti-neutrinos from supernova SN1987A was used in the past to set the model
independent limit $m_{\bar \nu_e}<30\,$eV \cite{Schramm:1987ra}.  This limit can
become significantly tighter under some specific assumptions. For example a
recent detailed reanalysis obtained, within the SN delayed explosion scenario,
the limit $m_{\bar \nu_e}<5.7\,$eV \cite{Loredo:2001rx}.

Since SN1987A, several efforts have been carried out to improve the
sensitivity of the method, while awaiting for the next explosion within our
Galaxy.  Often, these approaches rely on ``timing'' events related to the
collapse of the star core, that are used as benchmarks for measuring the
neutrino delays. The emission of gravitational waves \cite{gravit1,gravit2},
the $\nu_e$ neutronization burst \cite{gravit2}, the initial steep raise of the
neutrino luminosity \cite{steep}, and the abrupt interruption of the neutrino
signal due to a further collapse into a black hole \cite{blackhole} have been
used to this aim.  However, there are some drawbacks to these methods: firstly
only neutrinos with arrival time close to the benchmarks are used, and they
represent only a small fraction of the total; secondly the observation of the
benchmark events is not always certain, and in any case some model dependence
on the details of the SN explosion is generally introduced.

The method we want to propose is free from these drawbacks: it relies only on
the measurement of the neutrinos energies and arrival times, and it uses the
full statistics of the detected signal, thus allowing to extract the maximum
of information. It is also remarkably independent of particular astrophysical
assumptions, since no use is made of benchmarks events.  The basic idea is the
following: in the idealized case of vanishing experimental errors in the
determination of the neutrino energies and arrival times, and assuming an
arbitrarily large statistics and a perfectly black-body neutrino spectrum, one
could use the events with energy above some suitable value $E^*$ (to suppress
the mass effects) to reconstruct very precisely the evolution in time of the
neutrino flux and spectrum.  Once the time dependence of the full signal is
pinned down, the only parameter left to reconcile the time distribution of the
low energy neutrinos with the high energy part of the signal would be the
neutrino mass, that could then be nailed to its true value.  Of course, none
of the previous conditions is actually fulfilled.  In water \v{C}erenkov
detectors as SuperKamiokande (SK) the uncertainty on relative
timings is negligible; however, the errors in the energy
measurements are important and must be properly taken in to account. The
statistics is large but finite, and this not only represents a source of
statistical uncertainty, but also implies an upper limit on the useful values
of $E^*$.  Finally, the SN neutrino spectrum is not
perfectly thermal \cite{neutrino-spectrum}.  Nevertheless, a good sensitivity
to the mass survives and, as we will show, it will be possible to disentangle
with a good confidence the two cases $m_\nu =0$ and 1 eV. 


\smallskip

To test quantitatively the idea outlined above we proceed in two steps:

\smallskip
\noindent
{\it i)} First we generate a set of synthetic neutrino
signals, according to some suitable SN model. Neutrinos are then
propagated from the SN to the detector assuming two different SN-Earth
distances (10 and 20 kpc) and two different mass values $m_\nu = 0\,$
and $ 1\,$eV.  Finally, two different energy thresholds (5 and 10 MeV)
are used for the detection.  The result consists of several neutrino
samples, that hopefully would not differ too much from a real SN
signal.

\smallskip
\noindent
{\it ii)} The signals are then analyzed with the main goal of disentangling
with sufficient confidence the two cases $m_\nu = 0\,$ and $1\,$eV.  Only the
SN-Earth distance is assumed to be known (this information could be obtained
directly from optical observations or, with an uncertainty of the order of
25\%, from a comparison of the measured total energy with a theoretical
estimation of the binding energy released \cite{located}).  Other
quantities, like the spectral functions and the details of the time evolution
of the neutrino flux, are inferred directly from the data.


\smallskip

{\it Generation of the neutrino samples.}\ The time evolution of the
neutrino luminosity and average energy can be obtained from the
results of SN explosions simulations
\cite{burrows,woosley,totani,lieben}.  In the present analysis we use
the results of Woosley et al. \cite{woosley}.  This model is
characterized by a rather hard neutrino spectrum ($\langle E_{\bar
\nu_e}\rangle\gsim 20\,$MeV) that results in a large number of
detected events (8,800 in SK).  Still this is a conservative choice,
since it also implies a depletion in the number of low energy
neutrinos with a corresponding loss of informations on the neutrino
mass.  Inside the protoneutron star neutrinos are in thermal
equilibrium, and therefore it is reasonable to assume that the gross
features of their energy spectrum after emission can still be
described by a Fermi-Dirac distribution:
\begin{equation} \label{spectrum} 
\widehat S\left(E_\nu; \hat T(t),\hat \eta\right) = \frac{E_\nu^2/\hat T^3(t)} 
{F_2(0,\hat \eta)\, \left(1 + e^{E_\nu/\hat T(t) - \hat\eta(t)}\right)},  
\end{equation}
where $F_2(0,\hat \eta)$ is defined in (\ref{Fermiint}) and a `pinching' factor
$\hat\eta(t)$ has been introduced to simulate spectral distortions
\cite{neutrino-spectrum}.  From now on, quantities with a hat ($\hat\eta$, $\hat
T$, \dots) represent inputs to the Monte Carlo (MC), while quantities without
a hat will refer to results of data fitting.  We define the Fermi functions
and generalized Fermi integrals as:
\beqa 
f_n(x, \eta ) &=& \frac{x^n}{1 + e^{x - \eta }} \label{Fermifun} \\ 
F_n(y, \eta ) &=& \int^\infty_{y} {f_n(x, \eta)}\,dx \,. \label{Fermiint}
\eeqa
Using the time evolution of the average neutrino energy
$\overline{E_\nu}(t)$ as given in fig.~3 of ref. \cite{woosley} and
taking for simplicity $\hat\eta=3$ \cite{neutrino-spectrum} constant,
we compute the effective temperature that, together with $\hat\eta$, 
determines the neutrino spectrum at the source:
\beq \label{hattemp} 
\hat T(t) =  k_{3} \> \overline{E_\nu}(t) \,,
\eeq
where $k_{3} \equiv F_2(0,3)/F_3(0,3)\approx 4$.  Given that the
average energy is only mildly dependent on time \cite{woosley}, we
model the  evolution of the neutrino flux $\widehat \Phi(t)$
simply by taking it proportional to the luminosity as given in fig.~2
of ref. \cite{woosley}.

In the SK detector, $\bar\nu_e$'s are detected through the inverse $\beta$
decay reaction $\bar \nu_e\, p\to e^+\, n $ that has the energy threshold
$E_{\rm react}= m_e + \Delta m_{np} \simeq 1.8\,$MeV, where $m_e$ is the
electron mass and $\Delta m_{np} \simeq 1.3\,$MeV is the neutron to proton
mass difference.  The effect of the detection cross section $\sigma(E_\nu)$
\cite{Strumia:2003zx} is taken into account from the beginning by using the
function $\widehat S\big(E_\nu;\hat T(t),\hat\eta\big)\times \widehat
\Phi(t)\times \sigma(E_\nu)$ in the MC generator.  Each SN neutrino is labeled
by its emission time $t_\nu$ and by its energy $E_{\nu}$.  The corresponding
detected positron is also identified by an energy/time pair of values $(E^e,
t^e)$.  We generate $E^e$ according to a Gaussian distribution with central
value $E^e=E_{\nu} - \Delta m_{np}$ and variance $\sigma=0.15\, \sqrt{10\,{\rm
    MeV}/E^e}\,$ that corresponds to the SK energy resolution
\cite{SKresolution}.  Since the time resolution of the SK detector is very
precise, no error is assigned to $t^e$. For massive neutrinos, we redefine
$t^e=t_\nu+\Delta t_\nu$ by including the appropriate delay $\Delta t_\nu =
L\, m^2_\nu/2\,E^2_\nu $. A fixed number of 8,800 energy/time pairs $(E^e_i,
t^e_i)$ is generated in each run. This corresponds to the expected number of
$\bar \nu_e$ interactions within the SK fiducial volume for the spectrum of
the model in \cite{woosley} and a SN at $10\,$kpc. For $L=20\,$kpc this number
is resampled down by a factor of four.  Next, positrons with energies below
the detection threshold ($E_{\rm tr}=5\,$ or $10\,$Mev) are discarded, and as
a last step the origin of the time axis is set in coincidence with the first
positron detected ($t^e_1=0$) and each $t^e_i$ is accordingly redefined.  For
each set of values ($E_{\rm tr}\,,L\,,m_\nu$) forty different samples are
generated in this way.


\smallskip


{\it Analysis of the neutrino signal.} \ We analyze 
the neutrino signals by means of the likelihood function
\beq
{\cal L} =  S\big(\epsilon;T(t),\eta(t)\big) \times \Phi(t+\delta t;b,d,f)\times
\sigma(\epsilon) \,.
\eeq
The energy $ \epsilon = E^e+\Delta m_{np}$ of each neutrino is
inferred from the positron energy.  The
spectral function $S$ is assumed of the form (\ref{spectrum}) with a
time dependent effective temperature $T(t)$ and pinching factor
$\eta(t)$ fitted from the data.
 $\Phi(t+\delta t;b,d,f)$ describes the
time evolution of the neutrino flux, and the four parameters $\delta
t\,,b\,,d$ and $f$ account for its location on the time axis and detailed
shape. Finally, $\sigma(\epsilon)$ is the neutrino cross section.

%
\begin{figure}[t]
\vskip-2mm
\hskip-6mm
\epsfysize=60mm
\epsfxsize=85mm
 \epsfbox{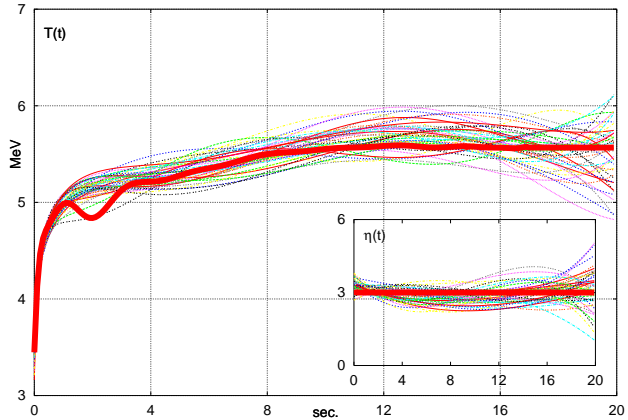}
\vskip-4mm
\caption[]{\baselineskip 12pt Comparison of the Monte Carlo
spectral temperature $\hat T(t)$ and pinching parameter $\hat \eta$
(thick lines), with $T(t)$ and $\eta(t)$ fitted for 40 
samples ($L=10\,$kpc, $m_\nu =0\,$).}
\label{1a}
\end{figure} 
%
%
%


{\it The spectral functions.}\ The two spectral functions $T(t)$ and
$\eta(t)$ are determined by fitting the first and second momentum of
the energy distribution to the mean energy $\overline{E_\nu}$ and mean
squared energy $\overline {E_\nu^2}$ of the {\it incoming} neutrino
flux (prior to detection).  Given a set of $n$ neutrinos with {\it
measured} energies $(\epsilon_1,\epsilon_2,\dots.\epsilon_n)$, and
assuming for the cross section the approximate form $\sigma(\epsilon)
\propto {\epsilon}^2$, we have:
\beq
\overline{E_\nu}  \approx  \frac{\sum_i \epsilon_i^{-1}}{\sum_i
  \epsilon_i^{-2}};  
\qquad \quad
\overline{E^2_\nu}  \approx  \frac{n}{\sum_i \epsilon_i^{-2}}. \label{emednu}
\eeq
From (\ref{delay}) we see that the typical neutrino delays are at most of the
order of few milliseconds. This is much shorter than the time scale over which
sizable variations of the spectral functions are likely to occur
\cite{burrows,woosley,totani,lieben}. Therefore, by binning over sufficiently
large time windows we can ensure that the determination of $T$ and $\eta$ for
each bin will not be affected by the delays. We use a set of 20 windows of
size increasing from 50 ms to 2 s, distributed over a signal duration 20 s.
After this time the neutrino flux is assumed to be too low to provide
additional informations.  To reduce further the possible effects of mass
induced time delays, only neutrinos with energy larger than $E^* = 15\,$MeV
are used.  For each time window [$t_{a}$,\,$t_{a+1}$] we perform a least
square fit to the following quantities:
\beqa \overline{E_\nu} (T,\eta)
&=&\frac{F_3(y,\eta)}{F_2(y,\eta)}\,T\,,
\label{emed} \\ 
\frac{\overline {E_\nu^2} (T,\eta)} {\left[\overline{E_\nu}
(T,\eta)\right]^2} &=& \frac{F_2(y,\eta)\> F_4(y,\eta)
}{F^2_3(y,\eta)}\,, \label{e2med} \eeqa
where the functions $F_n(y,\eta)$ defined in (\ref{Fermiint}), depend
on the temperature through $y = {E^*/T}$.  This yields the `best'
values $T(\bar t_{a})$ and $\eta(\bar t_{a})$ at $\bar
t_a=(t_a+t_{a+1})/2$.  Finally, in order to obtain two smooth
continuous functions, the points $T(\bar t_a)$ and $\eta(\bar t_a)$
are interpolated with two half integer power polynomials $P_{T,\eta}
\sim \sum_l c_l t^{l/2} $ with $l=0,1,2,\dots 10$ for $T$ and
$l=0,1,3,5$ for $\eta$.  In fig.~\ref{1a} the functions $\hat T(t)$
and $\hat\eta=3$ used in the MC (thick lines) are compared with forty
fits to neutrino samples generated with $m_\nu =0$ and $L=10\,$kpc.
These results remain  unchanged for $m_\nu =1\,$eV, while
for $L=20\,$kpc, due to the reduced statistics, the fits show a
somewhat wider dispersion.



{\it The neutrino flux.}\ Even if the details of the neutrino flux
evolution with time are not known, its gross features can be predicted
on rather solid theoretical grounds.  It is expected that a sharp
exponential rise, with a time scale of tenths of milliseconds, is 
followed by a power law decay, with a time scale of several
seconds. We model this behavior by means of a parametric analytical function
in which, roughly speaking, two parameters describe the rising and decaying
rates, and a third one accounts for the transition point:
\beqa \label{lum}
\Phi(t;b,d,f) \sim \frac{ e^{-f t^{-m}}}{(1 + b\, t^n)^{d}} 
\to  \casos{ll}
  e^{-f t^{-m}} &  (t \to 0)   \\
  t^{-n d} & (t \to \infty)\,.
\fincasos   \quad
\eeqa
This function can reproduce reasonably well the results of different models
\cite{woosley,totani,burrows,lieben}.  The two exponents $m$ and $n$ are fixed
to suitable integer values by means of a preliminary rough fitting procedure
to the data, and then are held constant throughout the analysis of all the
samples. For the model in \cite{woosley} we use $n=8$ and $m=2$.  Finally,
since the origin of times was arbitrarily set in coincidence with the
detection of the first neutrino while $\Phi$ vanishes at $t=0$, a fourth
parameter $\delta t$ is required to let $\Phi(t+\delta t)$ freely shift along the time
axis.


  
\begin{figure}[t]
\vskip-6mm
\hskip-6mm
\epsfysize=60mm
\epsfxsize=87mm
 \epsfbox{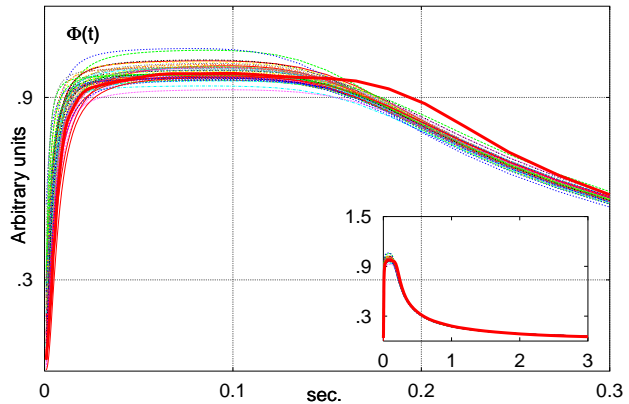}
\vskip-4mm
\caption[]{\baselineskip 12pt Comparison between the Monte Carlo flux
function $\widehat \Phi(t)$ (thick line), and the result of the
likelihood analysis for 40 samples ($L=10\,$kpc, $E_{tr}=5\,$MeV,
$m_\nu =0$). }
\label{1b}
\end{figure}

{\it The likelihood analysis.}\ The minimization of the negative
log-likelihood is carried out by means of the package MINUIT \cite{minuit}.
To avoid double minimums, we minimize with respect to the square of the
neutrino mass.  Given a value of $m^2_\nu$ the time delay of each neutrino is
computed according to its energy $\epsilon_i$, and subtracted from its arrival time
$t_i$.  The log-likelihood for the new array of times is then evaluated, and
minimized with respect to the other parameters $b\,,d\,,f$ and $\delta t$.  This
proceeds until the absolute minimum is found in the full five dimensional
parameter space.  There is a subtlety related to the cases when, especially
for large test masses, a neutrino is migrated to an early time where $\Phi=0$.
The problem is not just a numerical one of logarithm overflow. Due to the
uncertainty in the energy measurement, the firsts neutrinos detected can end
up in such a position without necessarily implying that the corresponding
distribution has vanishing probability. To account for this, for the relevant
neutrinos the energy uncertainty is converted into an uncertainty in the new
time position, and the corresponding contribution to the likelihood is
evaluated by convolving ${\cal L}(t)$ in (\ref{lum}) with a Gaussian of the
appropriated width.  In fig. \ref{1b} the results of the fitted fluxes for 40
samples with $L=10\,$kpc, $E_{\rm tr} = 5\,$MeV and $m_\nu =0\,$eV are compared
with the function $\widehat \Phi(t)$ used in the MC generator (thick line).

%
%


\smallskip 

\begin{figure}[t]
\vskip-4mm
\hskip-6mm
\epsfysize=60mm
\epsfxsize=87mm
 \epsfbox{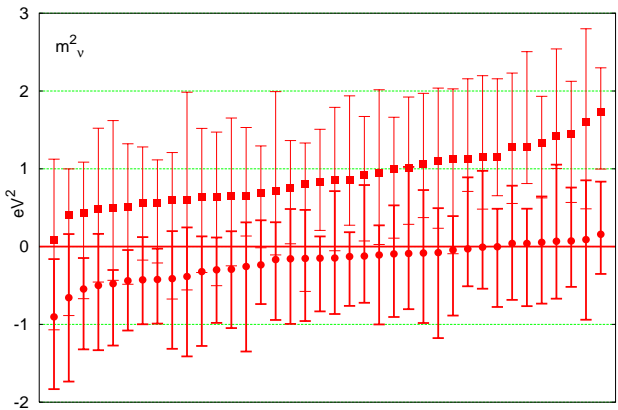}
\vskip-5mm
\caption[]{\baselineskip 12pt 
Fitted values of the mass and 95\% c.l. error bars for two sets of 40
analysis with $L=10\,$kpc and $E_{tr}=5\,$MeV, for $\hat m_\nu=0$
(circles) and $\hat m_\nu=1\,$eV (squares).  }
\label{2a}
\end{figure}

{\it Results.}\ Our results are summarized in fig.~\ref{2a} and in table~1.
Fig. \ref{2a} depicts the results of the mass fits for two sets of 40 samples
with $E_{\rm tr}=5\,$MeV, $L=10\,$kpc, $\hat m_\nu=0$ (circles) and $\hat
m_\nu=1\,$eV (squares).  The data points have been ordered with increasing value
of the best fit mass to put in evidence the difference between the two cases.
Error bars correspond to 95\% c.l., where   
the lower (upper) limit is computed by integrating the likelihood
from $m^2_\nu=\infty$ ($-\infty$) until reaching the 95\% of the area, while minimizing
with respect to the other parameters.  We have checked that the limits do not
change much if the integration is restricted to the physical region $m_\nu^2>0$.

 Table~1 summarizes some
results of the mass fitting. For each set of parameters ($E_{\rm
  tr},\,L,\,\hat m_\nu$) we have analyzed 40 samples.  The first raw gives the
percentage of times in which the 95\% c.l. lower limit $m^l_\nu$ is {\it
  larger} than the input mass $\hat m_\nu$.  The second raw refers to the
cases when the upper limit $m^u_\nu$ is {\it smaller} than $\hat m_\nu$.
Number in parentheses correspond to $L=20\,$kpc.  These figures characterize
the percentage of `failures' of the method, that therefore appears to be
reliable in about 90\%-95\% of the cases.  The third row gives the percentage
of times when $m_\nu=0\,$ is excluded at 95\% c.l. when the signal is
generated with $\hat m_\nu=1\,$eV. This characterizes the power of the method
for excluding a massless neutrino.  We see that in the most favorable case
($E_{\rm tr}=5\,$MeV, $L=10\,$kpc) the method is successful in more than 50\%
of the cases. The following three raws give the average over the
40 samples of the mass square best fit and of the 95\% c.l. lower and
upper limits (only for $L=10\,$kpc). These last figures are just
intended to give an idea of the quality of the fits (they would be
fully meaningful only if 40 SN could be observed).  

%
%
%

%
\begin{table}[t]  \renewcommand{\arraystretch}{.8}
\label{table}
\setlength{\tabcolsep}{.6mm}
\begin{tabular}{|c|c|c|c|c|}
\hline \vbox{\vskip8pt}\strut $E_{{\rm tr}}$
&\multicolumn{2}{c}{5\,MeV}\vline
&\multicolumn{2}{c}{10\,MeV}\vline\\[2pt] \hline \vbox{\vskip10pt}
$\hat{m}_\nu $ (eV)    & 0         & 1           & 0         & 1 \\ [4pt] \hline \vbox{\vskip10pt}
$m_\nu^l > \hat{m}_\nu\,$(\%)& 5 (11)& 4 \ (5) & 5 (10)& \ 11 (5)\\[2pt]
$m_\nu^u < \hat{m}_\nu\,$(\%)& 9 (6) & 2 \ (5) & 12 (6)& \ 10 (7) \\  [2pt]
$ m_\nu^l > 0\>$ \  (\%)        & --        & 55 (40) & --        & 28 (23)\\ [2pt] \hline
 \vbox{\vskip12pt} 
$\left<m_\nu^2\right>$ & --$0.1\pm0.4$ & $1.0\pm0.5$ & --$0.2\pm0.8$ & $\ 0.8\pm1.0$\\[2pt] 
 $\left<(m_\nu^l)^2\right>$& --$0.8\pm0.4$ & $0.1\pm0.6$ & --$1.6\pm1.0$ & --$0.8\pm1.2$\\[2pt]
 $\left<(m_\nu^u)^2\right>$ & $\ 0.6\pm0.5$ & $1.8\pm0.5$ & $\ 1.1\pm1.0$& $\ 2.2\pm1.1$\\[3pt] \hline
\end{tabular}
\caption{ \baselineskip 12pt 
Results of the analysis for the two neutrino masses $\hat m_\nu=0\,$ and
$1\,$eV, two energy thresholds of $5$ and $10\,$MeV and for two SN distances
of $10\,$kpc and, in parenthesis, $20\,$kpc.  The first (second) row give the
percentage of times in which the 95\% c.l. lower (upper) limit $m_\nu^l$
($m_\nu^u$) is larger (smaller) than $\hat m_\nu$.  The third row lists how
many times $m_\nu=0$ can be excluded (95\% c.l.) when $\hat m_\nu =1\,$eV.
The following three rows give, for $L=10\,$kpc, the mass
square best fit and the 95\% lower and upper limits averaged over 40
samples.}
\end{table}

From the results in the table it is apparent that low energy neutrinos are
crucial for the sensitivity of the method, and therefore a low detection
threshold is very important.  With $E_{\rm tr}=5\,$MeV the massless case is
excluded in about 50\% of the cases, while this drops to 25\% when $E_{\rm
  tr}=10\,$MeV.  Also, with the higher energy treshold the fluctuations of the
results over the 40 runs is doubled (last three rows).  Unfortunately, there
could be a dangerous background in the energy range between 5 and 10 MeV,
represented by photons originating from neutral current reactions off $^{16}$O
mainly produced by the more energetic $\mu$ and $\tau$ neutrinos \cite{oxygen}.
Softer neutrino spectra, like for example the ones predicted by the model of
Totani et al. \cite{totani}, would have the double benefit of drastically
reducing this background, while at the same time raising the number of $\bar
\nu_e$ emitted in the 5-10 MeV energy range from about 7\% of the present
analysis to about 20\%. In this case a better sensitivity to the mass could be
expected.  We should also mention that very recently it has been suggested
that water \v{C}erencov detectors could be modified  to allow tagging
of the inverse $\beta$ decay neutrons \cite{Beacom:2003nk}. This would 
eliminate the background from neutral current reactions and allow for 
lower thresholds.


{\it Numerical spectrum and neutrino oscillations.}\ The previous results have
been derived relying on two main simplifying approximations for the neutrinos
energy spectrum: {\it 1)}~the neutrino energies have been generated assuming
the `pinched' Fermi-Dirac spectrum given in eq. (\ref{spectrum}) and fitted,
as was described above, with a similar two-parameters energy distribution;
{\it 2)}~no effects of the neutrino oscillations were included in the
analysis.  In order to evaluate to what extent the sensitivity of the method
could be affected by these approximations, we have run a set of simulations in
which the neutrino energies were generated according to the shapes of the
numerical spectra given by Janka and Hillebrandt in \cite{neutrino-spectrum}.
A time dependent energy rescaling of the spectral shapes was introduced to
reproduce properly the time evolution of the mean energies as given in
\cite{woosley}.  We stress that since a two parameters Fermi-Dirac
distribution fits rather accurately the spectra obtained from the numerical
simulations, dropping our first simplification does not affect sensibly the
numerical results.

Neutrino oscillations can produce a composite spectrum corresponding to an
admixture of the original $\bar \nu_e$ spectrum with a harder component due to
$\bar \nu_x$ ($x=\mu,\tau$) \cite{Lunardini:2003eh}. Clearly, the resulting spectral
distortions will depend in first place on the size of the $\bar \nu_e-\bar \nu_x$
spectral differences. While it is often stated that these differences could be
quite sizable, and could yield up to a factor of two hierarchy between the
$\bar \nu_x$ and $\bar \nu_e$ average energies
\cite{burrows,woosley,totani,lieben}, recent and more complete analysis of SN
neutrino spectra formation \cite{KeilRaffeltJanka} indicate that this is not
the case: the inclusion of important interaction rates that were neglected in
previous works yields spectral differences that are only of the order of 10\%
\cite{KeilRaffeltJanka}.  At this level, identifying the two components of a
mixed spectrum would be a difficult task, and could represent a real challenge
for the study of SN neutrino oscillations in the $\bar \nu_e-\bar \nu_x$ channel.
However, for what concerns our analysis, this ensures that the results
discussed above are not affected much by neutrino oscillations.  To keep on the
safe side, we have tested the sensitivity of our method to oscillation effects
by running a set of simulations using the results of Woosley {\ea}
\cite{woosley} for which, as a consequence of negelcting important neutrino
reactions \cite{KeilRaffeltJanka}, the spectral differences are extreme
($\overline {E}_{\bar \nu_x} \approx 2\, \overline {E}_{\bar \nu_e}$). As we will show,
even in this (probably unrealistic) case, we find that the loss in sensitivity
to the neutrino mass is small. We can conclude that the effects of neutrino
oscillations do not endanger the applicability of our method and neither its
overall sensitivity to the neutrino mass.

As is discussed in \cite{Lunardini:2003eh}, depending on the type of the
neutrino mass hierarchy (normal or inverted) and on the size of $\sin^2
\theta_{13}$ (large $\gsim 10^{-4}$ or very small $\lsim 10^{-6}$), neutrino
oscillations could {\it i)} harden the $\bar \nu_e$ spectrum through a complete
spectral swap with $\bar \nu_\mu$ (inverted hierarchy, large $\theta_{13}$) {\it ii)}
mix the $\bar \nu_e$ spectrum with a fraction of about $\sin^2\theta_{12}\sim 1/4$ of
harder neutrinos (in the other cases). (The region $\sin^2\theta_{13}\sim
10^{-6}-10^{-4}$ would produce spectra with an intermediate amount of mixing.)
The first case can be studied without modifications in our procedure.
Clearly, a different spectrum would imply somewhat different numerical
results; however, this is analogous to the unavoidable uncertainty related to
the choice of the particular SN model since, for example, the $\bar \nu_e$ mean
energies that have been used in the present analysis \cite{woosley} are quite
close to the $\bar \nu_\mu$ mean energies of the model of Totani et al.
\cite{totani} .  The second case is more interesting since, for large spectral
differences, fitting a composite spectrum with just one effective spectral
temperature and one pinching parameter could degrade somewhat the sensitivity.

%
%
\begin{figure}[t]
\vskip-4mm
\hskip-6mm
\epsfysize=60mm
\epsfxsize=87mm
 \epsfbox{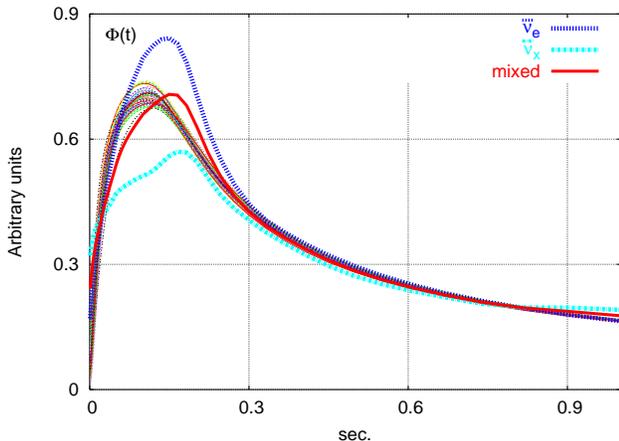}
\vskip-5mm
\caption[]{\baselineskip 12pt 
  Comparison between the $\bar \nu_e$ (upper thick line) and $\bar \nu_\mu$ (lower
  thick line) fluxes used in the Monte Carlo, the resulting $\bar \nu_e$ flux at
  the detection point (middle thick line), and the maximum likelihood fits for
  40 neutrino samples with a mixed composite spectrum. }
\label{mix-flux}
\vskip-5mm
\end{figure}
%
%
%
\begin{figure}[b]
\vskip-2mm
\hskip-6mm
\epsfysize=60mm
\epsfxsize=87mm
 \epsfbox{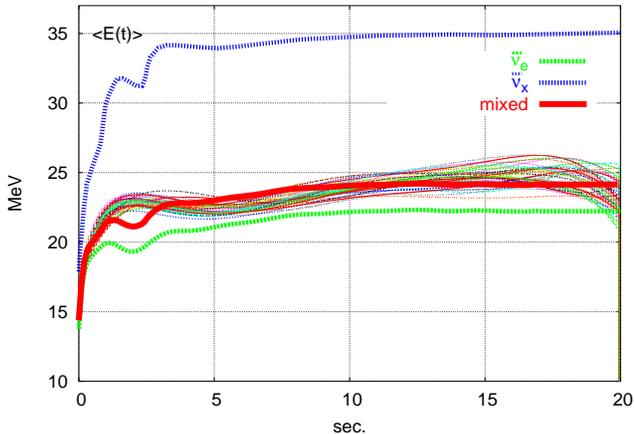}
\vskip-5mm
\caption[]{\baselineskip 12pt 
  Comparison between the $\bar \nu_e$ (lower thick line) and $\bar \nu_\mu$ (upper
  thick line) mean energies of the numerical spectra used in the Monte Carlo,
  the mean energy of the mixed composite spectrum (middle thick line), and the
  fitted values of $\overline E_{\bar \nu_e}(t)$ for 40 neutrino samples.}
\label{mix-energy}
\end{figure} 
%

We have carried out an analysis of forty neutrino samples with a mixed
composite spectrum as would result from a normal mass hierarchy, $\sin^2
\theta_{12} \approx 1/4$ and large $\theta_{13}$.  The original $\bar \nu_e$ and $\bar \nu_x$
fluxes taken from \cite{woosley} are depicted in fig. \ref{mix-flux} and
compared with the composite $\bar \nu_e$ flux at the detection point, as well as
with our maximum likelihood fits. The relative normalization of the fluxes was
computed assuming flavor equipartition of the integrated luminosities and the
time dependent average energies given in \cite{woosley}.  A similar comparison
between the mean energies of the two spectral components, the average energy
of the mixed spectrum and the average energies obtained from Fermi-Dirac
distributions with fitted spectral parameters $\hat T(t)$ and $\hat \eta(t)$ is
depicted in fig.  \ref{mix-energy}.

\begin{figure}[t]
\vskip-4mm
\hskip-6mm
\epsfysize=60mm
\epsfxsize=87mm
 \epsfbox{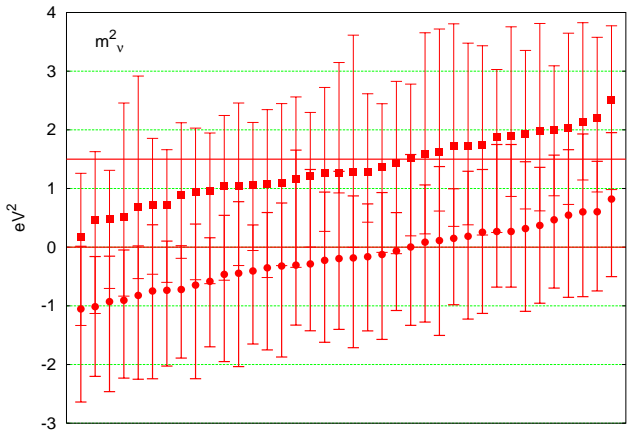}
\vskip-5mm
\caption[]{\baselineskip 12pt 
Fitted values of the mass and 95\% c.l. error bars for a set of 40
neutrino samples with mixed composite spectra  ($L=10\,$kpc, 
$E_{tr}=5\,$MeV,  $\hat m_\nu=0$ (circles), $\hat m_\nu=1.2\,$eV (squares)).}
\label{mix-band}
\end{figure}

For the mass fits we have taken $E_{\rm tr}=5\,$MeV and $L=10\,$kpc.  Since we
are essentially interested in the loss of sensitivity to the neutrino mass
with respect to the non-oscillation case, we have been searching for a mass
value that can reproduce results similar to those of the band plot in
fig.~\ref{2a} (namely, for a signal generated with $\hat m_\nu\neq 0$ we require
the massless case to be excluded at 95\% c.l. in at least 50\% of the runs).
As is shown in fig.~\ref{mix-band} our requirement is fullfilled for $\hat
m_\nu=1.2\,$eV, to be compared with $\hat m_\nu=1.0\,$eV for the non-oscillation
case.  We conclude that even in the unrealistic case of extremely large
spectral differences between the components of a mixed spectrum, fitting the
data with a single two parameters Fermi-Dirac distribution does not degrade
much the numerical results for the mass estimate. Given that the most recent
results suggest that the $\bar \nu_e$ and $\bar \nu_\mu$ spectra are in fact not
very different \cite{KeilRaffeltJanka}, the use of more complicated bimodal
energy distributions to improve the mass fits is probably not justified.

%
%
%


Before concluding, a few remarks are in order.  In this work we have not
carried out any deep study aimed to optimize the single fits to each different
neutrino sample (windows size, interpolating functions $P_{T,\eta}$, specific flux
function parameters).  A large number of different sets of data have been
analized through the very same procedure, in order to collect enough
informations about the sensitivity of the method in a reasonable amount of
time.  It is clear to us that optimizing the overall procedure in order to 
analze a specific sample (as would certainly be the case with a signal from a
real SN) can improve the sensitivity and shrink somewhat the uncertainties on
$m^2_\nu\,$.

Besides the effects of neutrino oscillations that were briefly analyzed in the
last paragraph, a few other issues could deserve further investigation.  For
example, assuming that the SK signal could be combined with negligible
uncertainty on the absolute timing with the signal detected at KamLAND, the
much better energy resolution and the lower threshold of this last detector
could enhance the sensitivity to the neutrino mass.  It also remains to see
what sensitivity could be achieved with a statistics one order of magnitude
larger, as would be available with the megaton neutrino detectors presently
under study \cite{HyperK,UNO,TITAND}.

\smallskip

 \centerline{\bf Acknowledgments.} 
\smallskip 
\noindent
We acknowledge conversations with H.-T. Janka, G. Raffelt and in particular
with A. Yu. Smirnov. J.I.Z. acknowledges hospitality from the Abdus Salam ICTP
in Trieste (Italy) during the final stage of this research, and Colciencias
for a scholarship for Doctoral Sudies. This work was supported in part by
Colciencias in Colombia under contract 1115-05-13809.



\begin{thebibliography}{99}

\bibitem{solar}
S.~Fukuda {\it et al.}  
Phys.\ Rev.\ Lett.\  {\bf 86},  5651 (2001);  
Q.~Ahmad {\it et~al.}  
Phys. Rev. Lett.  {\bf 87}, 071301 (2001). 

\bibitem{atmospheric}
Y.~Fukuda {\it et al.} 
Phys.\ Rev.\ Lett.\  {\bf 81} 1562 (1998); ibid.  {\bf 82}, 2644 (1999); 
M. Ambrosio {\it et al.} 
Phys.\ Lett.\ {\bf B434}, 451 (1998); ibid. {\bf B478}, 5 (2000).

\bibitem{kamland}
K. Eguchi {\it et al.} 
Phys. Rev. Lett. 90, 021802 (2003). 

\bibitem{absolutenumass}
H.~Pas and T.~J.~Weiler,
Phys.\ Rev.\ D {\bf 63}, 113015 (2001);
S.~Bilenky {\it et al.}, 
Phys.\ Rept.\  {\bf 379}, 69 (2003);

\bibitem{tritiumlimit}
J.~Bonn {\it et al.}, 
Prog. Part. Nucl. Phys. {\bf 48}, 133 (2002); 
V.~M.~Lobashev {\it et al.},  
Nucl. Phys. Proc. Suppl.\ {\bf 91}, 280 (2001).

\bibitem{doublebeta}
H.V. Klapdor-Kleingrothaus \ea, 
\epja{12}, 147 (2001); 
C.E. Aalseth \ea, 
Phys. Rev. D65,  092007 (2002).



\bibitem{WMAP}
D.~N.~Spergel {\it et al.}, 
Astrophys.\ J.\ Suppl.\ {\bf 148}, 175 (2003).



\bibitem{Colless:2001gk}
M.~Colless {\it et al.}, 
MNRAS 328, 1039 (2001).


\bibitem{Hannestad} 
S.~Hannestad,
JCAP {\bf 0305}, 004 (2003). 
O.~Elgaroy and O.~Lahav,
JCAP {\bf 0304}, 004 (2003). 


\bibitem{Zatsepin:1968}
G.T. Zatsepin, JETP Lett. 8 (1968) 205, 
[Zh. Eksp. Teor. Fiz. 8, 333 (1968)]; 
S.~Pakvasa and K.~Tennakone, Phys. Rev. Lett. 28,  1415 (1972).

\bibitem{Schramm:1987ra}
D.~N.~Schramm,
Comments Nucl.\ Part.\ Phys.\  {\bf 17}, 239 (1987).

\bibitem{Loredo:2001rx}
T.~J.~Loredo and D.~Q.~Lamb,
Phys.\ Rev.\ D {\bf 65}, 063002 (2002). 

\bibitem{gravit1}
D. Fargion, Lett. Nuovo Cim.  31, 499 (1981).

\bibitem{gravit2} 
N. Arnaud \ea, Phys. Rev. D 65, 033010 (2002).

\bibitem{steep}
T. Totani, Phys. Rev. Lett.  80, 2039 (1998).

\bibitem{blackhole}
J. F. Beacom, R. N. Boyd and A. Mezzacappa, Phys. Rev. Lett. 85, 3568 (2000);
Phys. Rev. D 63, 073011 (2001).


\bibitem{neutrino-spectrum}
H. -T. Janka and W. Hillebrandt,
Astron. Astrophys. {\bf 224}, 49 (1989).


\bibitem{located}
J.~F.~Beacom and P.~Vogel,
Phys.\ Rev.\ D {\bf 60}, 033007 (1999).  


\bibitem{burrows}
A.~Burrows, D.~Klein and R.~Gandhi,
Phys.\ Rev.\ D {\bf 45}, 3361 (1992).

\bibitem{woosley} 
S. E. Woosley \ea, 
         Astrophys. J.  433, 229 (1994).

\bibitem{totani} T. Totani \ea, 
Astrophys. J.  496, 216 (1998). 


\bibitem{lieben}
M.~Liebendorfer \ea, 
Phys.\ Rev.\ D {\bf 63}, 103004 (2001).


\bibitem{Strumia:2003zx}
A.~Strumia and F.~Vissani,
Phys.\ Lett.\ B {\bf 564}, 42 (2003).


\bibitem{SKresolution}
M.~Nakahata \ea,  
Nucl.\ Instrum.\ Meth.\ A {\bf 421}, 113 (1999).  



\bibitem{minuit}
F.~James and M.~Roos,
Comput.\ Phys.\ Commun.\  {\bf 10}, 343 (1975).


\bibitem{oxygen} 
K.~Langanke, P.~Vogel and E.~Kolbe,
Phys.\ Rev.\ Lett.\  {\bf 76}, 2629 (1996). 


\bibitem{Beacom:2003nk}
J.~F.~Beacom and M.~R.~Vagins,
arXiv:hep-ph/0309300.



\bibitem{Lunardini:2003eh}
See e.g. C.~Lunardini and A.~Y.~Smirnov,
JCAP {\bf 0306}, 009 (2003), 
and references therein. 

\bibitem{KeilRaffeltJanka} 
M.~T.~Keil, G.~G.~Raffelt and H.~T.~Janka,
Astrophys.\ J.\  {\bf 590}, 971 (2003);  
G.~G.~Raffelt, M.~T.~Keil, R.~Buras, H.~T.~Janka and M.~Rampp,
arXiv:astro-ph/0303226; 
M.~T.~Keil,
arXiv:astro-ph/0308228.


\bibitem{HyperK}
K.~Nakamura, 
talk given at the conference ``Neutrinos and Implications for Physics  
Beyond the Standard Model";  Suny, Stony Brook, October 11-13, 2002.
[www.physics.sunysb.edu/itp/conf/ neutrino/talks/nakamura.pdf]

\bibitem{UNO} 
C.~K. Jung, arXiv:hep-ex/0005046.  

\bibitem{TITAND}
 Y.~Suzuki \ea, 
 Presented at 2nd Workshop on Neutrino Oscillations and Their Origin (NOON
 2000), Tokyo, Japan, 6-18 Dec 2000.   
Published in ``Tokyo 2000, Neutrino oscillations and their origin'' 
 288-296 [arXiv:hep-ex/0110005]. 

\end{thebibliography}
\end{document}